\documentclass[12pt]{article}
\usepackage{amsmath}

\usepackage[super,compress]{cite}
\usepackage{graphicx}
\usepackage[polutonikogreek,english]{babel}
\usepackage{comment}
\renewcommand{\d}{\mathrm{d}}
\newcommand{\bpart}{{\mbox{\boldmath$\partial$}}}
\newcommand{\rlambda}{{\rm \textgreek{l}}}
\newcommand{\rnu}{{\rm \textgreek{n}}}
\newcommand{\on}{{\overline{n}}}
\newcommand{\otheta}{\overline{\vartheta}}
\newcommand{\Tr}{\mathop{\mathrm{Tr}}}
\newcommand{\Det}{\mathop{\mathrm{Det}}}
\newcommand{\bp}{{\mbox{\boldmath$p$}}}
\newcommand{\sgn}{\mathop{\mathrm{sgn}}\nolimits}
\newcommand{\oDelta}{{\overline{\Delta}}}

\allowdisplaybreaks[4]

\begin{document}

\title{Soft synchronous gauge in the perturbative gravity}

\author{V.M. Khatsymovsky \\
 {\em Budker Institute of Nuclear Physics} \\ {\em of Siberian Branch Russian Academy of Sciences} \\ {\em
 Novosibirsk,
 630090,
 Russia}
\\ {\em E-mail address: khatsym@gmail.com}}
\date{}
\maketitle

\begin{abstract}

An attempt to directly use the synchronous gauge ($g_{0 \lambda} = - \delta_{0 \lambda}$) in perturbative gravity leads to a singularity at $p_0 = 0$ in the graviton propagator. This is similar to the singularity in the propagator for Yang-Mills fields $A^a_\lambda$ in the temporal gauge ($A^a_0 = 0$). There the singularity was softened, obtaining this gauge as the limit at $\varepsilon \to 0$ of the gauge $n^\lambda A^a_\lambda = 0$, $n^\lambda = (1, - \varepsilon (\partial^j \partial_j )^{- 1} \partial^k ) $. Then the singularities at $p_0 = 0$ are replaced by negative powers of $p_0 \pm i \varepsilon$, and thus we bypass these poles in a certain way.

Now consider a similar condition on $n^\lambda g_{\lambda \mu}$ in perturbative gravity, which becomes the synchronous gauge at $\varepsilon \to 0$. Unlike the Yang-Mills case, the contribution of the Faddeev-Popov ghosts to the effective action is nonzero, and we calculate it. In this calculation, an intermediate regularization is needed, and we assume the discrete structure of the theory at short distances for that. The effect of this contribution is to change the functional integral measure or, for example, to add non-pole terms to the propagator. This contribution vanishes at $\varepsilon \to 0$. Thus, we effectively have the synchronous gauge with the resolved singularities at $p_0 = 0$, where only the physical components $g_{j k}$ are active and there is no need to calculate the ghost contribution.

\end{abstract}


MSC classes: 83C45; 83C47

keywords: general relativity; Feynman diagrams; synchronous gauge; functional integral; temporal gauge

\section{Introduction}

Despite the formal non-renormalizability of the general relativity (GR) at the perturbative level \cite{hooft}, it is possible to view it as an effective low-energy theory \cite{don,don1,don2,don3}, whose predictions do not depend on the details of the high-energy behavior of the underlying theory. Such an area of application could be, for example, long distance quantum corrections to Newton's potential; those caused by graviton exchange have been studied in a number of works \cite{don,don1,don2,don3,muz,akh,hamliu1,kk,kk1}.

In perturbative calculations, the synchronous gauge $g_{0 \lambda} = - \delta_{0 \lambda}$ (in notation in which the metric has the spacelike signature $(-,+,+,+)$) may be of interest. This is an analogue of the gauge $nA^a = 0$ in Yang-Mills theories, where $n$ is a constant 4-vector, mainly the temporal $A^a_0 = 0$ or axial $A^a_3 = 0$ gauge, used for quite some time \cite{Schw}. An attractive feature of this gauge is that there is no ghost field contribution. Besides, certain field components are eliminated from consideration. In gravity, by fixing four components of the metric tensor, the synchronous frame explicitly leaves us with six physically significant metric variables - the spatial metric. But the disadvantage of such a gauge is the appearance of a (double) pole at $p_0 = 0$, $( n p )^{- 2}$, in the longitudinal part of the gauge field propagator. This singularity is closely related to the fact that the gauge is not completely fixed by the condition $nA^a = 0$ and arbitrary time-independent gauge transformations are possible. To overcome this disadvantage, Landshoff's $\alpha$-prescription\cite{Land} ($1/p_0^2 \Rightarrow 1/(p_0^2 + \alpha^2), \alpha \to 0$) can be used. A justification for this prescription was given by Steiner\cite{Ste} by means of modifying ("softening") the gauge condition $nA^a = 0$, for example, $A^a_0 = 0$, as follows:
\begin{equation}                                                            
A^a_0 - \varepsilon ( \bpart^2 )^{- 1} \partial^j A^a_j = 0 , \quad j = 1, 2, 3,
\end{equation}

\noindent that is, making $n^\lambda$ a differential nonlocal operator. Then the nonphysical poles shift to the plain of complex momentum, $( n p )^{- 1} = (p_0 + i \varepsilon )^{- 1}$, $( \on p )^{- 1} = (p_0 - i \varepsilon )^{- 1}$.

In the soft synchronous gauge in gravity, $n^\lambda g_{\lambda \mu} = 0$, the contribution of the ghost field to the effective action turns out to have no pole. This non-pole contribution to the effective action turns out to be imaginary and, in fact, this means some factor in the functional integral measure. But we are just interested in the functional measure and show here that this factor tends to 1 (that is, the measure is not modified due to the ghost field) in the limit $\varepsilon \to 0$. We get an analogue of Landshoff's prescription.

The interest to the soft synchronous gauge and its influence on the functional measure due to ghosts may also be connected with a certain specifying the underlying theory as discrete, whereas the continuum GR is what we observe at low energies or large distances. In this regard, the simplicial approach, started with the Regge calculus \cite{Regge}, may be of interest, allowing one to formulate discrete gravity and extract from it predictions for physical effects/constants \cite{RocWilPL,RocWil,HamWil1}. The required symmetries are expected to be restored by taking into account all possible simplicial structures in the functional integral. For the simplest periodic simplicial structure with a 4-cubic cell divided by diagonals into 24 4-simplices \cite{RocWilPL}, and for the Regge action in the leading order in metric variations, we obtain a finite difference form of the continuum Hilbert-Einstein action \cite{our2}. In the functional integral approach, a non-simple functional measure arises from functional integration over a discrete connection in the connection representation of the Regge action. The initial point of the perturbative expansion is found by finding the extremum of the action (as usual) and the maximum of the measure. This leads to finite non-zero typical elementary length scales of Regge spacetime, proportional to the Planck scale. Around this point, we can also develop the perturbative expansion itself (UV finite) \cite{our1,khat}, containing both discrete analogues of all ordinary continuum diagrams, practically reproducing the finite ones, and diagrams with new vertices. The functional measure is known exactly in the limit when the elementary length scale in the temporal direction is arbitrarily small. Let the elementary length scale in the spatial direction still be determined by maximizing the measure. Then, to obtain a well-defined perturbative expansion around such an initial point, we should prohibit length variations in the temporal direction (or variations of the discrete ADM lapse-shift functions \cite{ADM1}). This means just using the (discrete version of the) synchronous gauge. We consider the graviton propagator in such a gauge \cite{our1}, and it has the aforementioned disadvantage - the presence of nonphysical poles at $p_0 = 0$. The present paper considers the "soft" synchronous gauge for gravity to overcome this disadvantage.

In what follows, we calculate the graviton propagator in the soft synchronous gauge, determine the ghost action, and calculate the effective action arising from the functional integration over the ghost field. It turns out to be important that the effective ghost action be of order $O( \varepsilon^2 )$, which turns out to be the case; then we can correctly pass to the limit $\varepsilon \to 0$ and disregard this action (as discussed in the paragraph with eqs (\ref{eT>>1}), (\ref{eeT<<1})). We also note (in Section \ref{typical}), as an example of the use of the soft synchronous gauge propagator in calculations, the coincidence, up to non-pole terms, of this propagator with the propagator considered in [\citen{hooft}] in the Coulomb-like gauge, the pole part of which, as stated there, determines the absorptive part of the S-matrix in the calculations there.

\section{Graviton propagator}

The Einstein-Hilbert action in the Planck units takes the form
\begin{eqnarray}                                                            
S & = & \frac{1}{8} \int \d^4 x \sqrt{ - g } g_{\lambda \mu , \nu} g_{\rho \sigma , \tau} \left( 2 g^{\lambda \rho} g^{\mu \tau} g^{\nu \sigma} - g^{\lambda \rho} g^{\mu \sigma} g^{\nu \tau} - 2 g^{\lambda \tau} g^{\mu \nu} g^{\rho \sigma} \right. \nonumber \\ & & \left. + g^{\lambda \mu} g^{\rho \sigma} g^{\nu \tau} \right) .
\end{eqnarray}

\noindent Now we would like to add a source term and a gauge fixing term to the action. In the synchronous gauge $g_{0 \lambda} = \eta_{0 \lambda}$. Therefore, in the soft synchronous gauge we consider the effective action
\begin{eqnarray}\label{S[J]}                                                
S^\prime [ J ] & = & S - \int \d^4 x \left[ J^{\lambda \mu} w_{\lambda \mu } + \frac{\rlambda}{4} (n^\lambda w_{\lambda \mu}) \eta^{\mu \rho } (n^\nu w_{\nu \rho}) \right] , \quad g_{\lambda \mu} = \eta_{\lambda \mu} + w_{\lambda \mu} , \nonumber \\ \eta^{\lambda \mu } & = & {\rm diag} (-1, 1, 1, 1) .
\end{eqnarray}

\noindent Here
\begin{equation}\label{n=nu-e}                                              
n^\lambda = \rnu^\lambda - \varepsilon \frac{ \partial_{\! \! \perp}^\lambda}{ \partial_{\! \! \perp}^2 }, \quad \partial_{\! \! \perp \lambda} = \partial_\lambda - \rnu_\lambda \frac{\rnu \partial}{ \rnu^2 } , \quad \rnu^\lambda = (1, 0, 0, 0) , \quad \partial_{\! \! \perp \lambda} = (0, \bpart) .
\end{equation}

\noindent Although $\rnu^\lambda$ can be considered fixed as shown in (\ref{n=nu-e}), for the sake of generality, whenever possible, we treat $\rnu^\lambda$ as an arbitrary vector.

Notations: For the coefficient $\rlambda$ and the 4-vector $\rnu$, we use upright letters to distinguish them from the indices $\lambda$ and $\nu$, which are written in italics.

Instead of the gauge fixing term in (\ref{S[J]}), we could consider the following:
\begin{equation}\label{lambda-term}                                         
- \frac{\rlambda }{4 } \int \d^4 x ( l^\sigma \partial_\sigma n^\lambda g_{\lambda \mu}) \eta^{\mu \rho } ( l^\tau \partial_\tau n^\nu g_{\nu \rho}) .
\end{equation}

\noindent Here $l^\lambda$ is some auxiliary vector indicating a coordinate direction. (One can even take $l^\lambda = \rnu^\lambda$.) We can include $l \partial$ into a redefinition of $n$: $(l \partial) n \Rightarrow n$. In particular, for $\rlambda^{- 1} = 0$, propagator  (\ref{G}) is invariant with respect to the scaling of $n$ and remains the same (for $l p \neq 0$ in momentum space; to $l p = 0$ this extends by continuity). At $\rlambda^{- 1} = 0$, this gauge reads
\begin{equation}                                                            
l^\nu \partial_\nu n^\lambda g_{\lambda \mu} = 0 ,
\end{equation}

\noindent that is, $n^\lambda g_{\lambda \mu}$ does not change in the direction indicated by the vector $\Delta x^\lambda = l^\lambda$. To specify $n^\lambda g_{\lambda \mu}$, it is enough to additionally specify $n^\lambda g_{\lambda \mu}$ on some remote 3d hypersurface. This set of additional conditions has an infinitely smaller measure compared to the set of bulk conditions provided by including the \rlambda-term in the action, and the proposed \rlambda-term (\ref{lambda-term}) is sufficient to single out bulk gravitational degrees of freedom. Thus, we can write the \rlambda-term both in the chosen form (\ref{S[J]}), homogeneous with respect to $w_{\lambda \mu} = g_{\lambda \mu} - \eta_{\lambda \mu}$ and inhomogeneous with respect to $g_{\lambda \mu}$ and with a predetermined initial point of expansion $g^{ ( 0 ) }_{\lambda \mu}$ (here $g^{ ( 0 ) }_{\lambda \mu} = \eta_{\lambda \mu}$), and in the homogeneous form (\ref{lambda-term}), which does not use a predefined point $g^{ ( 0 ) }_{\lambda \mu}$.

Returning to the effective action $S^\prime [ J ]$, we will vary it with respect to $w_{\lambda \mu}$ in order to determine the graviton propagator. Equating the result of the variation to zero gives
\begin{eqnarray}\label{d2g=J+nf+d2g}                                        
\partial^2 w_{\lambda \mu} & = & 4 J_{\lambda \mu} + \rlambda \on_\mu n^\nu w_{\lambda \nu} + \rlambda \on_\lambda n^\nu w_{\mu \nu} + \partial_\mu \partial^\nu w_{\lambda \nu} + \partial_\lambda \partial^\nu w_{\mu \nu} \nonumber \\ & & - \eta^{\nu \rho} \partial_\lambda \partial_\mu w_{\nu \rho} - \eta_{\lambda \mu} \partial^\nu \partial^\rho w_{\nu \rho} + \eta_{\lambda \mu} \eta^{\nu \rho} \partial^2 w_{\nu \rho} .
\end{eqnarray}

\noindent We denote $f_\lambda = n^\mu w_{\lambda \mu}$. Taking the divergence of both sides using the operator $\partial^\mu ( \cdot )$, we get
\begin{equation}                                                            
\on_\mu \partial^\mu f_\lambda + \on_\lambda \partial^\mu f_\mu = - 4 \rlambda^{-1} \partial^\mu J_{\lambda \mu} .
\end{equation}

\noindent This gives $f_\lambda$,
\begin{equation}                                                            
n^\mu w_{\lambda \mu} = f_\lambda = - 4 \rlambda^{- 1} (\on \partial )^{- 1} \partial^\mu J_{\lambda \mu} + 2 \rlambda^{- 1} \on_\lambda (\on \partial )^{- 2} \partial^\mu \partial^\nu J_{\mu \nu} .
\end{equation}

\noindent Taking the trace of (\ref{d2g=J+nf+d2g}) gives $\eta^{\lambda \mu} \partial^2 w_{\lambda \mu} - \partial^\lambda \partial^\mu w_{\lambda \mu}$ in terms of $J_{\lambda \mu}$ and the found $f_\lambda$,
\begin{equation}                                                           
\eta^{\lambda \mu} \partial^2 w_{\lambda \mu} - \partial^\lambda \partial^\mu w_{\lambda \mu} = - 2 \eta^{\lambda \mu} J_{\lambda \mu} - \rlambda \on^\lambda f_\lambda .
\end{equation}

\noindent Multiplying (\ref{d2g=J+nf+d2g}) by $n^\mu$ allows to express $\partial^\mu w_{\lambda \mu} - \eta^{\mu \nu} \partial_\lambda w_{\mu \nu}$ in terms of $J_{\lambda \mu}$ and the found $f_\lambda$ and $\eta^{\lambda \mu} \partial^2 w_{\lambda \mu} - \partial^\lambda \partial^\mu w_{\lambda \mu}$,
\begin{eqnarray}                                                           
( n \partial ) ( \partial^\mu w_{\lambda \mu} - \eta^{\mu \nu} \partial_\lambda w_{\mu \nu} ) & = & - 4 n^\mu J_{\lambda \mu} + \partial^2 f_\lambda - \partial_\lambda \partial^\mu f_\mu - \rlambda ( n \on ) f_\lambda - \rlambda \on_\lambda n^\mu f_\mu \nonumber \\ & & - n_\lambda ( \eta^{\mu \nu} \partial^2 w_{\mu \nu} - \partial^\mu \partial^\nu w_{\mu \nu} ) .
\end{eqnarray}

\noindent In turn, multiplying the found $\partial^\mu w_{\lambda \mu} - \eta^{\mu \nu} \partial_\lambda w_{\mu \nu}$ by $n^\lambda$ and using the found $f_\lambda$ gives $\eta^{\lambda \mu} w_{\lambda \mu}$. Substituting the found $\eta^{\lambda \mu} w_{\lambda \mu}$ back into $\partial^\mu w_{\lambda \mu} - \eta^{\mu \nu} \partial_\lambda w_{\mu \nu}$, we find $\partial^\mu w_{\lambda \mu}$. Thus, the terms with $w_{\lambda \mu}$ appearing on the RHS of (\ref{d2g=J+nf+d2g}) have been found. Substituting them there, we obtain the propagator, $w_{\lambda \mu} = G_{\lambda \mu \sigma \tau} J^{\sigma \tau}$,
\begin{eqnarray}\label{G}                                                  
G_{\lambda \mu \sigma \tau} & = & - i \langle w_{\lambda \mu} w_{\sigma \tau} \rangle = \frac{2}{\partial^2 } [ L_{\lambda \sigma} ( n, \on ) L_{\mu \tau} ( n, \on ) + L_{\mu \sigma} ( n, \on ) L_{\lambda \tau} ( n, \on ) \nonumber \\ & & - L_{\lambda \mu} ( n, n ) L_{\sigma \tau} ( \on, \on ) ] - \frac{2}{\rlambda } \frac{\eta_{\lambda \sigma} \partial_\mu \partial_\tau + \eta_{\mu \tau} \partial_\lambda \partial_\sigma + \eta_{\mu \sigma} \partial_\lambda \partial_\tau + \eta_{\lambda \tau} \partial_\mu \partial_\sigma}{(n \partial)(\on \partial)} \nonumber \\ & & + \frac{2}{ \rlambda } \left[ \frac{\partial_\lambda \partial_\mu ( n_\sigma \partial_\tau + n_\tau \partial_\sigma )}{(n \partial)^2 (\on \partial)} + \frac{\partial_\sigma \partial_\tau ( \on_\lambda \partial_\mu + \on_\mu \partial_\lambda )}{(n \partial) (\on \partial)^2 } \right] \nonumber \\ & & - \frac{2}{ \rlambda } \frac{( n \on ) \partial_\lambda \partial_\mu \partial_\sigma \partial_\tau }{(n \partial)^2 (\on \partial)^2 } ,
\end{eqnarray}

\noindent where
\begin{equation}                                                           
L_{\lambda \mu} ( m, n ) \stackrel{\rm def }{=} \eta_{\lambda \mu} - \partial_\lambda \frac{m_\mu }{ m \partial } - \frac{n_\lambda }{ n \partial } \partial_\mu + \frac{(m n) \partial_\lambda \partial_\mu}{(m \partial )( n \partial )}
\end{equation}

\noindent for 4-vectors $m$, $n$. For
\begin{equation}                                                           
m^\lambda = \rnu^\lambda - \varepsilon_1 \frac{ \partial_{\! \! \perp}^\lambda}{ \partial_{\! \! \perp}^2 } , \quad n^\lambda = \rnu^\lambda - \varepsilon_2 \frac{ \partial_{\! \! \perp}^\lambda}{ \partial_{\! \! \perp}^2 }
\end{equation}

\noindent it takes the form
\begin{equation}\label{L=eta-nunu/nu^2-...}                                
L_{\lambda \mu} ( m, n ) = \eta_{\lambda \mu} - \frac{ \rnu_\lambda  \rnu_\mu }{ \rnu^2 } - \frac{ \partial_{\! \! \perp \lambda} \partial_{\! \! \perp \mu} }{ \partial_{\! \! \perp}^2 } + \frac{ \partial^2 }{ \partial_{\! \! \perp}^2 } \frac{ ( \rnu^2 \partial_{\! \! \perp \lambda} + \varepsilon_1 \rnu_\lambda ) ( \rnu^2 \partial_{\! \! \perp \mu} + \varepsilon_2 \rnu_\mu ) }{ \rnu^2 ( \rnu \partial - \varepsilon_1 ) ( \rnu \partial - \varepsilon_2 ) } ,
\end{equation}

\noindent where $\varepsilon_1 = \pm \varepsilon$, $\varepsilon_2 = \pm \varepsilon$ are of interest.

\section{Ghost contribution}

Now consider the corresponding Faddeev-Popov ghost field contribution, which should give the determinant of the operator ${\cal O}$ in the functional integral measure, defined under the infinitesimal gauge (coordinate) transformations $\delta x^\lambda = \xi^\lambda ( x )$,
\begin{equation}                                                           
\delta g_{\lambda \mu} = - g_{\lambda \nu } \partial_\mu \xi^\nu - g_{\mu \nu } \partial_\lambda \xi^\nu - \xi^\nu \partial_\nu g_{\lambda \mu } = - \xi_{\lambda ; \mu} - \xi_{\mu ; \lambda} = - \partial_\mu \xi_\lambda - \partial_\lambda \xi_\mu + 2 \Gamma^\nu_{\lambda \mu} \xi_\nu ,
\end{equation}

\noindent by the formula
\begin{equation}                                                           
\delta f_\lambda = n^\mu \delta g_{\lambda \mu} = - (n \partial ) \xi_\lambda - \partial_\lambda ( n \xi ) + 2 n^\mu \Gamma^\nu_{\lambda \mu} \xi_\nu \equiv - {\cal O}_\lambda{}^\mu \xi_\mu .
\end{equation}

\noindent The ghost Lagrangian density is $\otheta^\mu {\cal O}_\mu{}^\lambda \vartheta_\lambda$ with anticommuting vector fields $\vartheta_\lambda$, $\otheta^\lambda$, but we transform the expression for $\Det {\cal O}$.\footnote{Here and below, the symbol $\Det$ means a determinant of the operator sense, in contrast to $\det$, the determinant of a matrix.} To do this, we factor out from ${\cal O}$, $\Det {\cal O}$ their values ${\cal O}_{( 0 )}$, $\Det {\cal O}_{( 0 )}$ at $\Gamma^\nu_{\lambda \mu} = 0$. We have
\begin{equation}                                                           
{\cal O}_{( 0 ) \mu}^{- 1}{}^\lambda = ( n \partial )^{- 1} \delta^\lambda_\mu - \frac{1}{2} ( n \partial )^{- 2} \partial_\mu n^\lambda
\end{equation}

\noindent (the ghost propagator). Then
\begin{equation}                                                           
( {\cal O}_{( 0 )}^{- 1} {\cal O} )_\mu{}^\lambda = \delta^\lambda_\mu - 2 ( n \partial )^{- 1} n^\nu \Gamma^\lambda_{\mu \nu} + ( n \partial )^{- 2} \partial_\mu n^\nu n^\rho \Gamma^\lambda_{\nu \rho} .
\end{equation}

So far, the calculations are valid for $n^\lambda$, which is an arbitrary differential operator independent of $x$. Further, in some places a specific form of $n^\lambda$ is implied, in particular such that $n \partial = \rnu \partial - \varepsilon$.

Next, we apply the operation $ ( n \partial )^{- 1} \| g \|^{- 1} n \partial ( \cdot ) \| g \|$ to this, that is, we multiply ${\cal O}_{( 0 )}^{- 1} {\cal O}$ by $ ( n \partial )^{- 1} \| g \|^{- 1} n \partial $ on the left and by the metric matrix $ \| g \| $ on the right. This gives the operator
\begin{eqnarray}\label{d-1g-1dO0-1Og}                                      
& & [ ( n \partial )^{- 1} \| g \|^{- 1} n \partial {\cal O}_{( 0 )}^{- 1} {\cal O} \| g \| ]^\lambda{}_\mu \equiv (1 + M)^\lambda{}_\mu = \delta^\lambda_\mu + ( n \partial )^{ - 1 } g^{ \lambda \nu } n^\rho ( g_{\nu \rho , \mu } - g_{\mu \rho , \nu } ) \nonumber \\ & & + ( n \partial )^{ - 1 } g^{ \lambda \nu } ( \rnu^\rho - n^\rho ) g_{ \mu \nu , \rho } + ( n \partial )^{ - 1 } g^{ \lambda \nu } ( n \partial )^{ - 1 } \partial_\nu n^\rho n^\sigma \Gamma_{\mu , \rho \sigma } .
\end{eqnarray}

\noindent This operation has the determinant equal to one and leaves $\Det ( {\cal O}_{( 0 )}^{- 1} {\cal O} )$ unchanged. Here $M = 0$ at $\varepsilon = 0$ (the Faddeev-Popov ghosts decouple). Up to a normalization constant $\Det {\cal O}_{( 0 )}$, the determinant of interest $\Det {\cal O}$ is defined by
\begin{equation}                                                           
\ln \Det {\cal O} = \ln \Det (1 + M) = \Tr \ln (1 + M) = \Tr \left( M - \frac{1}{2} M^2 + \dots \right) = \Tr M .
\end{equation}

\begin{figure}[h]
	\centering
	\includegraphics[scale=1]{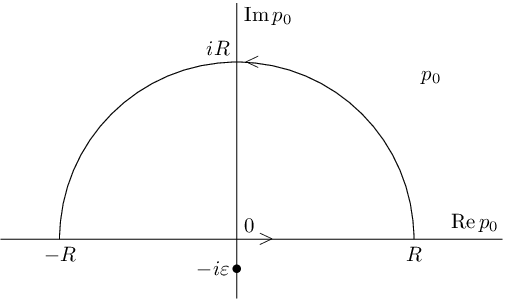}
	\caption{Integration contour in the plane of complex $p_0$.}
	\label{f1}
\end{figure}

\noindent Here $\Tr M^j$ at $j \geq 2$ in the momentum representation is a combination of the integrals $\int^{ + \infty }_{ - \infty } ( i p_0 - \varepsilon )^{-k} \d p_0 $ at $k \geq j$ which are equal to zero. This can also be illustrated by contour integration in the complex $p_0$ plane in Fig.~\ref{f1}.
There we consider the integral along a contour consisting of the original segment $[ - R , + R ]$, $R \to \infty$, and a semicircle $p_0 = R \exp ( i \phi )$, chosen between $\phi \in [0,  \pi ] $ and $\phi \in [2 \pi,  \pi ] $ so as not to cover the pole $p_0 = - i \varepsilon$. At $j \geq 2$, both the semicircle integral vanishes at $R \to \infty$ and the residue at the pole $p_0 = - i \varepsilon$, which defines the contour integral, is equal to zero. When $j = 1$, the integral over the semicircle does not vanish (the residue at the pole is still equal to zero by the construction of the contour). As a result, we are left with $\Tr M$,
\begin{eqnarray}\label{ghost}                                              
\Tr M & = & \Tr \left \{ \frac{ \varepsilon }{n \partial } g^{\lambda \mu} \frac{\partial_{\! \! \perp}^\nu }{\partial_{\! \! \perp}^2 } g_{\lambda \mu , \nu } + \frac{1 }{ n \partial } g^{\lambda \rho} \frac{1 }{ n \partial } \rnu_\rho \rnu \partial \left[ \frac{\rnu^\mu \rnu^\nu }{ \rnu^2 } - \varepsilon \frac{\rnu^\mu }{\rnu^2 } \frac{\partial_{\! \! \perp}^\nu }{\partial_{\! \! \perp}^2 } \right. \right. \nonumber \\ & & \left. \left. - \varepsilon \frac{\rnu^\nu }{\rnu^2 } \frac{\partial_{\! \! \perp}^\mu }{\partial_{\! \! \perp}^2 } + \frac{ \varepsilon^2 }{ \rnu^2 } \frac{ \partial_{ \! \! \perp }^\mu \partial_{ \! \! \perp }^\nu }{ ( \partial_{ \! \! \perp }^2 )^2 } \right] \Gamma_{ \lambda , \mu \nu } \right \} \nonumber \\ & = & \int \partial_k \ln ( - g ) \d^4 x \int \frac{ \d^4 p }{ ( 2 \pi )^4 } \frac{ \varepsilon }{ i \rnu ( p - q ) - \varepsilon } \cdot \frac{ i p^k}{ - \bp^2 } \nonumber \\ & & + \int \rnu_\lambda \Gamma^\lambda_{\mu \nu} \d^4 x \int \frac{ \d^4 p }{ ( 2 \pi )^4 } \frac{ i \rnu p }{[i \rnu ( p - q ) - \varepsilon] ( i \rnu p - \varepsilon ) } \left[ \frac{\rnu^\mu \rnu^\nu }{ \rnu^2 } + \varepsilon \frac{\rnu^\mu }{\rnu^2 } \frac{ i p_{\! \! \perp}^\nu }{p_{\! \! \perp}^2 } \right. \nonumber \\ & & \left. + \varepsilon \frac{\rnu^\nu }{\rnu^2 } \frac{ i p_{\! \! \perp}^\mu }{p_{\! \! \perp}^2 } - \frac{ \varepsilon^2 }{ \rnu^2 } \frac{ p_{ \! \! \perp }^\mu p_{ \! \! \perp }^\nu }{ ( p_{ \! \! \perp }^2 )^2 } \right] .
\end{eqnarray}

\begin{figure}[h]
	\centering
	\includegraphics[scale=1.5]{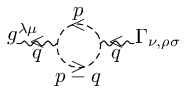}
	\caption{Effective diagram for the ghost contribution (\ref{ghost}).}
	\label{f2}
\end{figure}

\noindent In the first equality in the second term in braces, we take the projection of $\partial$ onto $\rnu$, $\partial_\lambda \to \partial_{\| \lambda } = \rnu \partial ( \rnu^2 )^{- 1} \rnu_\lambda$, as the only component that provides a nonzero result according to what was just said. We also have:
\begin{equation}                                                           
\int \frac{ \d p_0}{ 2 \pi } \frac{ 1 }{ i \rnu p - \varepsilon } = - \frac{1 }{ 2 ( - \rnu^2)^{1 / 2} } \sgn \varepsilon
\end{equation}

\noindent at an arbitrary scale of $\rnu$ and sign of $\varepsilon$. The two terms in braces in (\ref{ghost}) have the structure described by the diagram in Fig.~\ref{f2}
(in fact, this is exactly what appears in place of the tadpole diagram for the ghost Lagrangian density $\otheta^\mu {\cal O}_\mu{}^\lambda \vartheta_\lambda$ after transforming ${\cal O}$ into (\ref{d-1g-1dO0-1Og}); also here the internal lines denote not only the ghost propagator $(n \partial)^{- 1}$, but also the nonlocality $(\partial_{\! \! \perp}^2)^{- 1}$ from the ghost-graviton vertex). The first term turns out to vanish for any of two reasons: first, due to the antisymmetry of the integrand with respect to $p_{\! \! \perp}^\lambda$, and second (concerning the bulk contribution to it), because the integrand is a total derivative with respect to $x$. Again, due to the antisymmetry in $p_{\! \! \perp}^\lambda$, the part of the second term that is linear in $\varepsilon$ contributes zero. This term turns out to be independent of the momentum $q$ flowing through the diagram, and we obtain the integration over $\d^4 x$ of the pointwise product of $g^{\lambda \rho } \rnu_\rho$ and $\Gamma_{\lambda , \mu \nu} $ which is $ \rnu_\lambda \Gamma^\lambda_{\mu \nu}$.

The integrals over $\d^3 p_{\! \! \perp} = \d^3 \bp$ appearing in (\ref{ghost}) in their current form diverge at large momenta. As mentioned in Introduction, we consider the case when the underlying theory is discrete and in such a discrete formulation the integration in momentum space is carried out over the momentum $p_k$, varying within the limits $- \pi / b_{\rm s} < p_k < + \pi / b_{\rm s}$, where $b_{\rm s}$ is the lattice step along the spatial coordinate $\Delta x^k = b_{\rm s}$, the derivative $\partial_k$ or $i p_k$ is replaced by $b_{\rm s}^{ - 1 } \Delta_k$ with the finite difference $\Delta_k = \Delta_k ( p_k ) = \exp (i b_{\rm s} p_k ) - 1 $. The integrals of interest take the form
\begin{eqnarray}                                                           
& & \int \frac{ \d^3 \bp}{ ( 2 \pi )^3 } = \frac{1 }{ b_{\rm s}^3 }, \quad \int \frac{ \d^3 \bp}{ ( 2 \pi )^3 } \frac{ p^j p^k }{ ( \bp^2 )^2 } = \frac{ \delta^{ j k } }{3 } \int \frac{ \d^3 \bp}{ ( 2 \pi )^3 } \frac{ 1 }{ \bp^2 } \nonumber \\ & & \Rightarrow \frac{ \delta^{ j k } }{3 } \int \frac{ \d^3 \bp}{ ( 2 \pi )^3 } \frac{ b_{\rm s}^2 }{ \sum^3_{l = 1} \oDelta_l ( p_l ) \Delta_l ( p_l ) } = \frac{ \delta^{ j k } }{3 } \int \frac{ \d^3 \bp}{ ( 2 \pi )^3 } \frac{ b_{\rm s}^2 }{ \sum^3_{l = 1} [2 \sin (p_l b_{\rm s} / 2)]^2 } \nonumber \\ & & = \frac{1.05 \dots }{ 12 } \frac{ \delta^{j k}}{ b_{\rm s} },
\end{eqnarray}

\noindent where, for definiteness, the numerical coefficient in the last equality is specified on the basis of some table integral ([\citen{MagOber}], page 137).

Thus, for $\ln \Det {\cal O} = \Tr M$ or for the ghost contribution to the action $S_{\rm ghost}$ we have:
\begin{eqnarray}\label{ln-det-O}                                           
\ln \Det {\cal O} & = & - \frac{1}{2} \sgn \varepsilon \int \left( \frac{1 }{b_{\rm s}^3 } \Gamma^0_{0 0} - \frac{1.05 \dots }{ 12 } \frac{ \varepsilon^2 }{ b_{\rm s} } \delta^{j k} \Gamma^0_{j k} \right) \d^4 x , \nonumber \\ S_{\rm ghost} & = & - i \ln \Det {\cal O} = - \frac{1}{2} \int \rnu_\lambda \Gamma^\lambda_{\mu \nu} m^{\mu \nu} \d^4 x , \nonumber \\ m^{\mu \nu} & \equiv & \frac{i \sgn \varepsilon }{ ( - \rnu^2 )^{3 / 2} } \left[ \frac{ \rnu^\mu \rnu^\nu }{ b_{\rm s}^3 } - \frac{1.05 \dots }{ 12 } \frac{ \varepsilon^2 }{ b_{\rm s} } \left( \eta^{\mu \nu} - \frac{ \rnu^\mu \rnu^\nu }{ \rnu^2 } \right) \right] .
\end{eqnarray}

\noindent When $\varepsilon = 0$, the value of $\ln \Det {\cal O}$ is zero. But we aim to pass to this limit from an arbitrarily small but nonzero $\varepsilon$, at which the theory can be defined and allows to perform functional integral calculations. In the general case, $\ln \Det {\cal O}$ is real and does not seem to be bounded from below or (more importantly for the convergence of the functional integral) from above over the entire range of changes in the field variables. In the phenomenological definition of the functional integral, we proceed from the above 3D lattice with $N \times N \times N$ sites, $N$ being large, with the spatial coordinate lattice step $\Delta x^k = b_{\rm s}$. To determine the (improper) functional integration over the metric, especially if we have not an oscillating, but a monotonic exponential, we mean introducing the maximum length $l_{\rm max}$ (IR regularization), which, taking into account the lattice step $\Delta x^k = b_{\rm s}$, gives a certain upper bound on the absolute value of the metric components $g_{j k}$. We can also introduce a minimum 3-volume $V_{\rm min} = ( \det \| g_{j k} \| )^{1 / 2}_{\rm min} \Delta^3 x$ (UV regularization), which, for $\Delta^3 x$ fixed (as $b_{\rm s}^3$), means a lower limit on $\det \| g_{j k} \|$. Together with the upper bound on the absolute value of the metric components $g_{j k}$, this means an upper bound on the absolute value of the components $\gamma^{j k}$ (the reciprocal matrix to $g_{j k}$).

The result also contains integration over $\d x^0$ and derivative $\partial_0$. To provide an upper bound on the absolute value of the resulting $S_{\rm ghost}$, we also need to introduce a temporal coordinate lattice step $\Delta x^0 = b_{\rm t}$ and some large but finite size of the time integration interval $T$.

The Christoffel symbols appearing in (\ref{ln-det-O}) can be written as
\begin{eqnarray}\label{Gamma000Gamma0jj=}                                  
& & 2 \Gamma^0_{0 0} = \left [\ln ( g_{0 j} \gamma^{j k} g_{0 k} - g_{0 0} ) \right ]_{, 0} - \frac{ g_{0 j} \gamma^{j k}_{ , 0 } g_{0 k} + g_{0 0 , j } \gamma^{j k} g_{ 0 k } }{ - g_{ 0 0 } + g_{0 j} \gamma^{j k} g_{0 k} } , \nonumber \\ & & 2 \Gamma^0_{ j k } \delta^{ j k } = \left ( \frac{ g_{ j k } \delta^{j k}}{ - g_{ 0 0 } + g_{0 l} \gamma^{l m} g_{0 m} } \right )_{ , 0 } + \frac{( - g_{ 0 0 } + g_{0 l} \gamma^{l m} g_{0 m} )_{, 0}}{( - g_{ 0 0 } + g_{0 n} \gamma^{n p} g_{0 p} )^2} g_{j k} \delta^{ j k } \nonumber \\ & & \phantom{ 2 \delta^{ j k } \Gamma^0_{ j k } = } + \frac{ g_{0 l} \gamma^{l m} (2 g_{m j , k } - g_{j k , m}) - 2 g_{ 0 j , k }}{ - g_{ 0 0 } + g_{0 n} \gamma^{n p} g_{0 p} } \delta^{ j k } ,
\end{eqnarray}

\noindent where $g_{0 0}$, $g_{0 j}$ should be substituted according to the introduced gauge at $\rlambda^{- 1} = 0$:
\begin{equation}                                                           
g_{0 j} = \varepsilon \frac{\partial^k}{\bpart^2} g_{j k}, \quad g_{0 0} = - 1 + \varepsilon \frac{\partial^j}{\bpart^2} g_{0 j} = - 1 + \varepsilon^2 \frac{\partial^j \partial^k}{\left(\bpart^2 \right)^2} g_{j k} .
\end{equation}

\noindent It can be seen that $\Gamma^0_{0 0} \Rightarrow O( \varepsilon^2 )$, $\Gamma^0_{ j k } \delta^{ j k } \Rightarrow O( \varepsilon )$ (regarding the bulk contribution when integrating over $\d^4 x$, i.e. without total derivatives). The assumed intermediate IR and UV regularization ensures that for a sufficiently small $\varepsilon$, the exponent $\ln \Det {\cal O}$ is a bounded functional of the metric with the coefficient $\varepsilon^2$.

If we restrict the variation of $x^0$ to some interval of large but finite size $T$, as noted above, then the momentum $p_0$ is discrete with a step $\sim T^{- 1}$. Integrals over $\d p_0$ of functions having factors $(p_0 \pm i \varepsilon)^{- k}$, $k \geq 1$, are replaced by discrete sums. These sums tend to integrals in the limit $\varepsilon \to 0$, $T \to \infty$ only if the discretization step $\sim T^{- 1}$ is negligibly small compared to $\varepsilon$,
\begin{equation}\label{eT>>1}                                              
T^{- 1} \ll | \varepsilon |, ~~~ | \varepsilon | T \to \infty .
\end{equation}

\noindent The above considered upper bound on the absolute value of the resulting $\ln \Det {\cal O}$ should be proportional to $T$, and if in addition $\ln \Det {\cal O}$ were $O( \varepsilon )$, equation (\ref{eT>>1}) would not allow one to conclude that the ghost contribution to the effective functional measure is negligible. However, it is important that, as we have found, $\ln \Det {\cal O} = O( \varepsilon^2 )$, so to obtain the vanishing ghost contribution to the effective functional measure it suffices to require
\begin{equation}\label{eeT<<1}                                             
\varepsilon^2 T \to 0
\end{equation}

\noindent in the limit $\varepsilon \to 0$. Obviously, there are such variants of the dependence of $T \to \infty$ on $\varepsilon \to 0$, in which both (\ref{eT>>1}) and (\ref{eeT<<1}) are fulfilled, for example, $T \sim |\varepsilon|^{- 3 / 2}$.

Thus, we should pass to the limit $\varepsilon \to 0$ simultaneously with $T \to \infty$ in such a way that conditions (\ref{eT>>1}), (\ref{eeT<<1}) are satisfied. This gives $S_{\rm ghost} = 0$.

Strictly speaking, when applying a discrete regularization to $S_{\rm ghost}$, one should calculate $S_{\rm ghost}$ also in the discrete framework. We have verified that discretizing $x^0$ (then, in particular, $p_0 \in [ - \pi / b_{\rm t} , \pi / b_{\rm t} ]$) in addition to discretizing $x^k$, $k = 1, 2, 3$, leads to nonzero $\Tr M^n$, $n > 1$, which turn out to be proportional to powers of $b_{\rm t}$ and form a power series in $\varepsilon$. For a sufficiently small $\varepsilon$, this power series converges, and we come to the same conclusion that $S_{\rm ghost}$ is a bounded functional of the metric times $\varepsilon^2$.

Perturbatively, the expansion of $S_{\rm ghost}$ begins with terms that are bilinear in $w_{\lambda \mu}$,
\begin{eqnarray}                                                           
S_{\rm ghost} & = & - \frac{1}{2} \int \rnu_\lambda (\eta^{\lambda \rho} - w^{\lambda \rho} + O( w^2 ) ) \Gamma_{\rho , \mu \nu} m^{\mu \nu} \d^4 x \nonumber \\ & = & \frac{1}{4} \int \rnu_\lambda w^{\lambda \rho}  (2 w_{\rho \mu , \nu} - w_{\mu \nu , \rho}) m^{\mu \nu} \d^4 x + O( w^3 )
\end{eqnarray}

\noindent up to boundary terms. Then, in particular, the graviton propagator is modified, $G_{\lambda \mu \sigma \tau} \to \widetilde{G}_{\lambda \mu \sigma \tau}$. To find this modification, we can redefine the source term,
\begin{eqnarray}\label{w=G(J+Aw)}                                          
\widetilde{J}^{\lambda \mu} & = & J^{\lambda \mu} - \frac{\delta S_{\rm ghost}}{\delta  w_{\lambda \mu}} , \nonumber \\ w_{\lambda \mu} & = & \widetilde{G}_{\lambda \mu \sigma \tau} J^{\sigma \tau} = G_{\lambda \mu \sigma \tau} \widetilde{J}^{\sigma \tau} = G_{\lambda \mu \sigma \tau} \left( J^{\sigma \tau} + \frac{1}{2} m_{\! \! \perp}^{\sigma \nu} \partial_\nu \rnu^\rho w_{\rho \pi} \eta^{\pi \tau} \right. \nonumber \\ & & \left. - \frac{1}{2} \rnu^\sigma m_{\! \! \perp}^{\rho \nu} \partial_\nu w_{\rho \pi} \eta^{\pi \tau} + \frac{1}{4} \rnu^\sigma \partial^\tau m^{\nu \rho} w_{\nu \rho} - \frac{1}{4} m^{\sigma \tau} \rnu^\nu \partial^\rho w_{\nu \rho} \right) \nonumber \\ & \equiv & G_{\lambda \mu \sigma \tau} ( J^{\sigma \tau} + A^{\sigma \tau \nu \rho} w_{\nu \rho} ) ,
\end{eqnarray}

\noindent where
\begin{equation}                                                           
m_{\! \! \perp}^{\lambda \mu} = C \left( \eta^{\lambda \mu} - \frac{ \rnu^\lambda \rnu^\mu }{ \rnu^2 } \right), \quad m_{\! \! \perp}^{\lambda \mu} \partial_\mu = C \partial_{\! \! \perp}^\lambda , \quad C \equiv \frac{ - i \sgn \varepsilon }{ ( - \rnu^2 )^{3 / 2}} \frac{1.05 \dots}{12} \frac{ \varepsilon^2 }{ b_{\rm s }} .
\end{equation}

\noindent Symbolically (\ref{w=G(J+Aw)}) can be solved for $w_{\lambda \mu}$ as
\begin{equation}\label{w=GJ+GAGJ+...}                                      
w = G ( J + A w ) = G J + G A G J + G A G A G J + \dots .
\end{equation}

\noindent Here, the structure of the matrix $A \equiv A^{\sigma \tau \nu \rho} $ in relation to the indices $\sigma , \tau $ can be described as a combination of $m_{\! \! \perp}^{\sigma \pi} \partial_\pi$, $\rnu^\sigma$, $\rnu^\sigma \partial^\tau$, $m^{\sigma \tau}$. Using the projectors $L_{\lambda \mu}$ in the form (\ref{L=eta-nunu/nu^2-...}) in $G_{\nu \rho \sigma \tau}$ (\ref{G}) (at $\rlambda^{-1} = 0$), we can see that each of these monomials, contracted with $G_{ \dots \sigma \tau }$ (i. e. on the left), cancels the graviton poles of $G_{ \dots \sigma \tau }$ (of $(\partial^2)^{- 1}$). The structure of the matrix $A \equiv A^{\sigma \tau \nu \rho} $ w. r. t. the indices $\nu , \rho$ can be described as a combination of $\rnu^\rho$, $m_{\! \! \perp}^{\rho \pi} \partial_\pi$, $m^{\nu \rho}$, $\rnu^\nu \partial^\rho$. Again, each of these monomials, contracted with $G_{ \nu \rho \dots }$ (i. e. on the right), cancels the graviton poles of $G_{ \nu \rho \dots }$. As a result, beginning from the 2nd term in (\ref{w=GJ+GAGJ+...}), each $G$ is multiplied by $A$ on the right and/or left. Therefore, only the first term, i. e. the original graviton propagator $G_{\lambda \mu \sigma \tau}$ has a pole at $\partial^2 = 0$.

Each $\rnu$ in $A$ brings a smallness of order $O( \varepsilon )$, each $m$ in $A$ brings the smallness of order $O( \varepsilon^2 )$. Each term in $A$ contains $\rnu$ and $m$ and leads to a smallness of order $O( \varepsilon^3 )$. Note that the smallness of the ghost corrections is generally $O( \varepsilon^2 )$, as we discussed above; this proportionality to $\varepsilon^2$ is found not in the bilinear form, but in the trilinear (in $w_{\lambda \mu}$) term in $\Gamma^0_{0 0}$ (\ref{Gamma000Gamma0jj=}).

Thus, the ghost-modified graviton propagator $\widetilde{G}$ differs from the original one $G$ by non-pole terms of order $O( \varepsilon^3 )$.

\subsubsection{Example of using soft synchronous gauge propagator in calculation}\label{typical}

The Coulomb-like Prentki gauge was considered in gravity in [\citen{hooft}]. It was noted there that the Faddeev-Popov ghost propagator for this gauge has no pole, and in the considered (one-loop) calculations in gravity this led to the conclusion that the ghost field does not contribute to the absorptive part of the S-matrix. By the same argument, the absorptive part of the S-matrix remains the same when replacing the graviton propagator in the soft synchronous gauge $G_{\lambda \mu \sigma \tau}$ (\ref{G}) (at $\rlambda^{-1} = 0$), which has projectors $L_{\lambda \mu}$ (\ref{L=eta-nunu/nu^2-...}), with the Prentki propagator of [\citen{hooft}], since they differ in non-pole terms. The latter is due to the fact that both these propagators differ from the propagator with the projector $L^{(0)}_{\lambda \mu} = \eta_{\lambda \mu} - \rnu_\lambda  \rnu_\mu / \rnu^2 - \partial_{\! \! \perp \lambda} \partial_{\! \! \perp \mu} / \partial_{\! \! \perp}^2 $ (which appears to be the projector $\overline{\delta}_{\lambda \mu}$ from [\citen{hooft}]) by non-pole terms. So using the soft synchronous gauge propagator there may be by replacing it with the Prentki one.

\section{Conclusion}

The method of softening the temporal gauge in Yang-Mills theories also works in the case of the synchronous gauge in GR. The nonphysical poles at $p_0 = 0$ of the graviton propagator in the complex $p_0$ plane are shifted to the imaginary region.

An attractive feature of this gauge is that we can disregard the ghost field contribution in the limit $\varepsilon \to 0$. Besides, we are left with 6 physically significant variables - the spatial metric.

To define the ghost contribution to the effective action, we need a regularization. As such a regularization, we assume the discrete form of the theory itself. Taking into account the ghost contribution, the theory can be defined in the functional integral framework for nonzero $\varepsilon$, at least for sufficiently small $\varepsilon$, if it is regularized.

The important thing is that the ghost contribution to the effective action is proportional to $\varepsilon^2$: therefore we can pass to the limit $\varepsilon \to 0$ in a consistent manner, as discussed in the paragraph with equations (\ref{eT>>1}), (\ref{eeT<<1}).

Selecting a gauge means selecting a set of configurations over which the summation in the functional integral will be performed. Violation of gauge symmetry in the discrete case means that the calculated physical quantities depend on the gauge. (Although there is an invariant part that reproduces the diagrams of standard perturbation theory, as mentioned in the Introduction.) Taking all configurations into account in the functional integral or averaging over possible gauges restores symmetry. Previously, we considered \cite{khat} the de Donder type gauge and configurations on which the functional measure was not obtained as a closed expression, but a mathematical model was used. The synchronous gauge now considered corresponds to configurations on which the functional measure is known explicitly.

\section*{Acknowledgments}

The present work was supported by the Ministry of Education and Science of the Russian Federation.

\end{document}